\documentclass{article}
\usepackage{graphicx}
\usepackage{longtable}

\newcommand{\mm}{ \mu^+\mu^-}

\begin{document}

\pagestyle{plain}
\footskip 1.5 cm

\title{Rotating Metal Band Target for Pion
Production at Muon Colliders and Neutrino Factories}
\author{B.J. King$^{1,2}$, N.V. Mokhov$^3$, N. Simos$^2$ and R.J. Weggel$^2$}
\maketitle

\begin{abstract}

  A conceptual design is presented for a high power pion production
target for muon colliders that is based on a rotating metal band. Three
candidate materials are considered for the target band: Inconel alloy 718,
titanium alloy 6Al-4V (grade 5) and nickel.
% All three target materials look attractive and one may be preferred over the
%others depending on the specifications of the incident pulsed proton
%beam and also on further design evaluation.
A pulsed proton beam tangentially intercepts a chord of the target
band that is inside a 20 tesla tapered solenoidal magnetic pion
capture channel similar to designs
previously considered for muon colliders and neutrino factories.
 The target band has a radius
of 2.5 meters and is continuously rotated at
approximately 1 m/s to carry heat away from the production region and
into a water cooling tank.
The mechanical layout and cooling setup of the target are described,
including the procedure for the routine replacement of the target band.
A rectangular band cross section is assumed, optionally with I-beam
struts to enhance stiffness and minimize mechanical vibrations.
%For Inconel and nickel,
%the band is assumed to be 6 mm wide and 100 mm high and the proton
%beam is assumed to have an elliptical gaussian profile with horizontal
%and vertical r.m.s widths of ?? mm and ?? mm, respectively;
%the corresponding assumed dimensions for the titanium
%alloy band are: a band that is 20 mm wide and 100 mm high
%and a proton beam with r.m.s widths and heights of ?? m.
 Results are
presented from realistic MARS Monte Carlo computer simulations of the pion
yield and energy deposition in the target and from ANSYS finite element
calculations for the corresponding shock heating stresses.
%Either a rectangular or I-beam band cross section may be used, with
%the choice of geometry and dimensions depending on the target material
%and proton beam parameters.
The target scenario is predicted to perform satisfactorily and with
conservative safety margins for multi-MW pulsed proton beams.

\end{abstract}

\footnotetext[1]{Corresponding author, email: bking@bnl.gov }
\footnotetext[2]{Brookhaven National Laboratory, P.O. Box 5000,
Upton, NY 11973-5000}
\footnotetext[3]{Fermi National Accelerator Laboratory, P.O. Box 500,
Batavia, IL 60510-0500}

%[redo]The target band has a 2.5 m radius and has an I-beam cross section that
%is 6 cm high
%and with a 0.6 cm thick webbing. The pion capture scenario and
%proton beam parameters are as specified for the Study II base-line
%targetry option, i.e. capture into a 20 tesla tapered solenoidal
%channel with proton beam fills at 2.5 Hz containing 6 short bunches,
%each spaced by 20 milliseconds, of $1.67 \times 10^{13}$ 24 GeV protons.

\section{Introduction and Overview}
%%%%%%%%%%%%%%%%%%%%%%
\label{sec:intro}

\begin{figure}[t!] % fig 1
\centering
\includegraphics[width=5.0in]{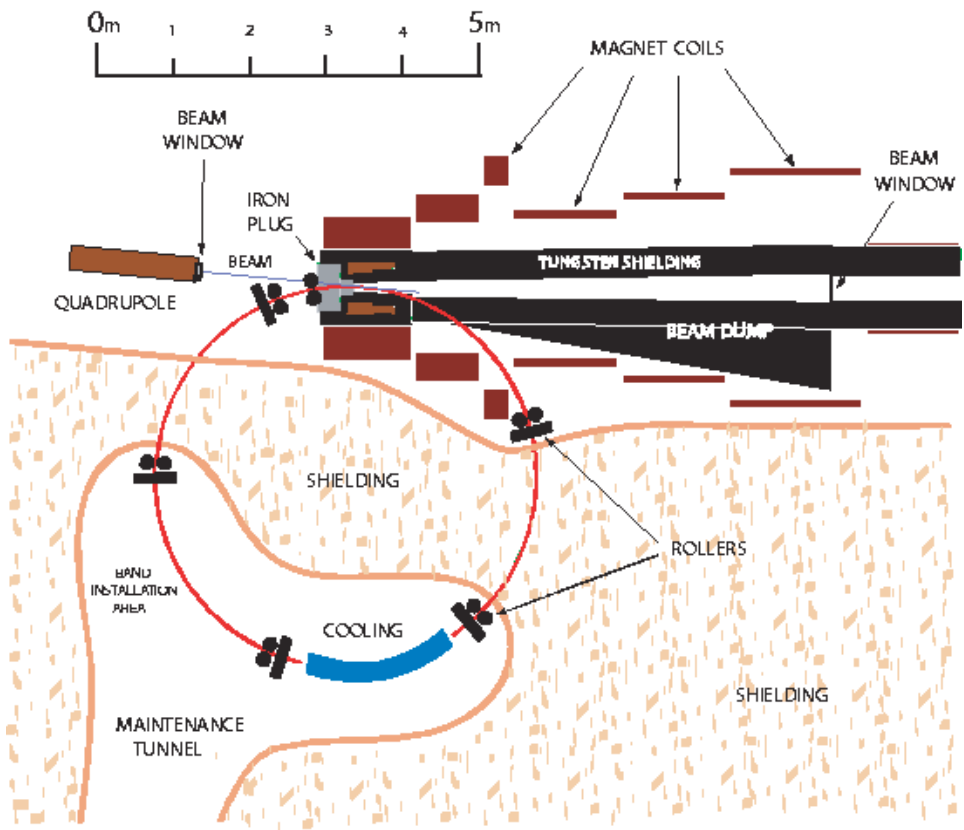}
\caption{A conceptual illustration of the targetry setup.}
\label{layout}
\end{figure}

\begin{figure}[t!] % fig 1
\centering
\includegraphics[width=4.0in]{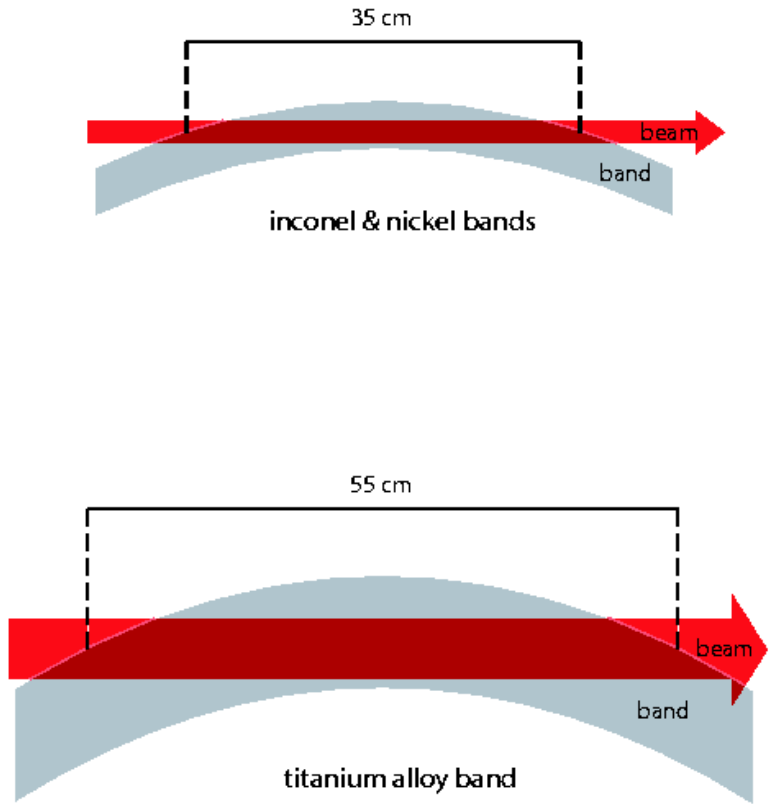}
\caption{Plan views of the
passage of the proton beam through the target bands, for
the Inconel or nickel band options (top) and for the titanium
alloy band option (bottom). The plots have the same scale and have
a vertical:horizontal
aspect ratio of approximately 5.4:1. The band curvature is the
same in both cases -- a 2.5 m radius of curvature -- but the
intersection length (55 cm) in the 20 mm thick titanium alloy band
is longer than in the 8 mm thick Inconel or nickel bands
(intersection lengths of 35 cm) because the intersection length
scales as the square root of the band thickness.
}
\label{band_and_beam_plan}
\end{figure}

\begin{figure}[t!] % fig 1
\centering
\includegraphics[height=3.0in]{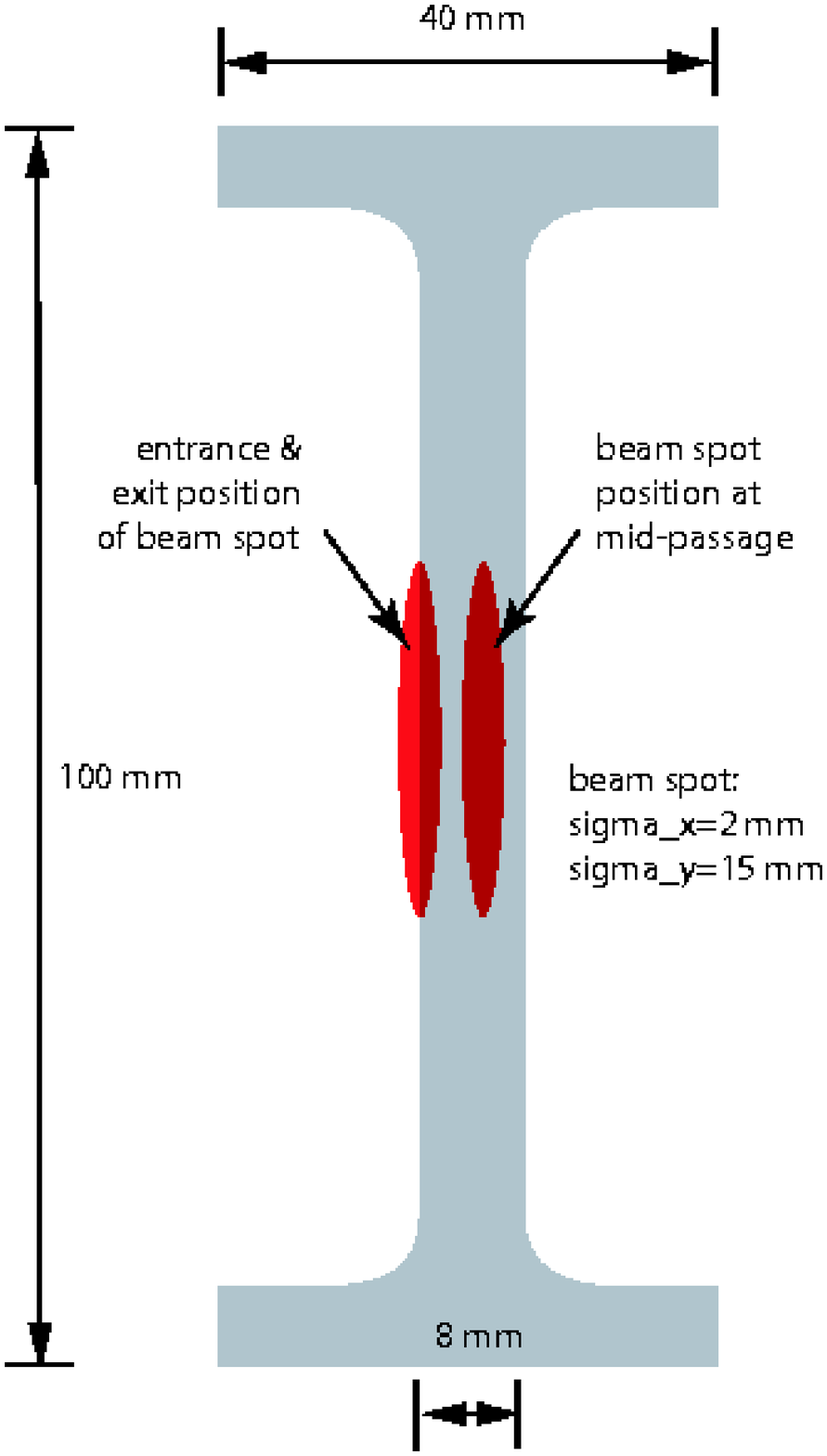}
\hspace{0.5in}
\includegraphics[height=3.0in]{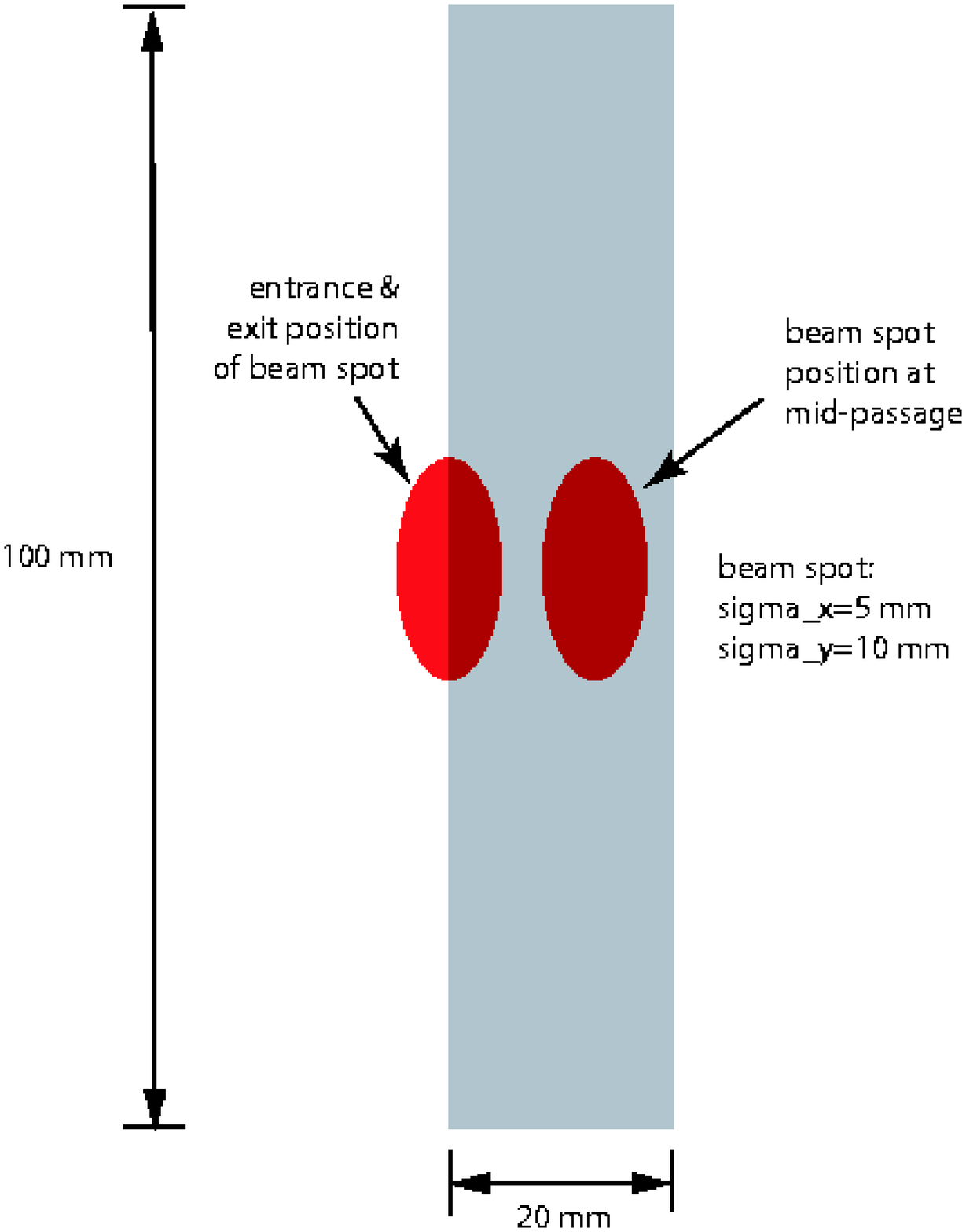}
\caption{
Cross-sectional views of the passage of the proton beam through the
target band, for the Inconel or nickel band options (left) and for
the titanium alloy band option (right). The horizontal position of
the beam spot in the band webbing varies along the interaction region
due to the curvature of the bands.}
\label{band_and_beam_xsec}
\end{figure}

\begin{table*}[htb!]
\caption{Specifications of the target band and assumed proton
beam parameters.}
\begin{tabular}{|r|ccc|}
\hline
Property       &  Inconel 718  &  Ti-alloy  & nickel \\ 
\hline
target band radius, [m]        &  2.5  & 2.5 & 2.5 \\
band thickness, [mm]           &   8   & 20 & 8 \\
band webbing height, [mm]      &   100   & 100 & 100 \\
full width of band flanges, [mm] &  40     & -- & 40 \\
beam path length in band, [cm]  &  35     & 55 & 35 \\
proton interaction lengths ($\lambda$)    & 2.1  & 2.0 & 2.3 \\
weight of band, [kg]       & 169  & 139 & 183 \\
horizontal beam-channel angle ($\alpha$), [mrad]  &  100 & 100 & 100 \\
rms beam spot size at target (horizontal), [mm]   & 2 & 5 & 2 \\
rms beam spot size at target (vertical), [mm]   & 15  & 10 & 15 \\
\hline
\end{tabular}
\label{target_band_specs}
\end{table*}

 The design of a pion production target for a muon collider or neutrino factory
is challenging because of the combination of high average power and large instantaneous
energy depositions from the pulsed proton beam, the geometric constraints
from the capture solenoid surrounding the target, and the desire to maximize
the pion yield through use of transversely
thin targets constructed from elements with
high, or at least medium, atomic numbers.

  Other target options that have previously been
considered for either muon colliders or neutrino factories
include liquid mercury jets~\cite{book96,status,study2}
and a radiation
cooled graphite rod~\cite{study1}. This paper presents a solid-target
option that is based on a rotating band geometry. Similar
conceptual designs for rotating band targets have been presented
previously~\cite{study2,PAC99_band_target,nufact99_band_target,RAL_band_target,nufactbandnote}.

%An abbreviated and less complete version of this paper is included
%as an appendix in the study II report~\cite{study2}.

%..point to layout figure; partially schematic [use NIM abstract]
  A plan view of the targetry setup for the band target option is shown
in figure~\ref{layout}. A 2.5 meter radius circular target band threads
through a solenoidal magnetic capture channel to tangentially intercept
the proton beam. Three metals are considered as candidates for the target
band: Inconel alloy 718, titanium alloy 6Al-4V grade 5 and pure nickel.
The pion capture channel is a slight modification of a previously
presented conceptual design~\cite{status,study2}, as will be discussed
further in section~\ref{sec:channel}.

  The proton beam enters the center of the target band webbing at a
glancing angle, and the beam
center traverses approximately two
interaction lengths of target material before
the surviving protons exit the target
due to the curvature of the band.
The cross sectional dimensions of the band and its orientation relative to
the proton beam are shown in figures~\ref{band_and_beam_plan}
and~\ref{band_and_beam_xsec}; the specifications of the band
and the proton beam dimensions are enumerated in table~\ref{target_band_specs}.
Inconel and nickel were studied for identical
band dimensions and proton beam parameters, whereas the titanium alloy band
was thicker, with no I-beam flanges required for stiffness, and used a
larger proton beam spot.
The circulating band is cooled by passage through a water
tank located in a separate radiation-shielded maintenance enclosure.

  The sections in this paper discuss, in order: 1) the range of
expected proton beam parameters, 2) the properties of the candidate
target materials and the specifications of the target band, 3) the
drive and support
rollers for the target band, 4) considerations for operating the
target region in an air environment, 5) required modifications to
the pion capture and decay channel in order to incorporate the
rotating band, 6) cooling of the band in a water tank, 7) radiation
damage and the replacement scheme for the target band, 8) MARS
Monte Carlo simulations of pion yield and the beam energy
deposition distribution, 9) beam-induced shock heating stresses
on the target band, and 10) overall conclusions on the
rotating band target design scenario.

\section{Incident Proton Beam Specifications}
%%%%%%%%%%%%%%%%%%%%%%%%%%%%%%%%%%%%%%%%%%%%%
\label{sec:protons}

  Pion sources for muon colliders have similar requirements to those
for the related technology of neutrino factories. However, there is
a greater emphasis on high charge proton bunches because, for high
luminosity muon collider parameters, the produced pion cloud
must eventually be transformed into muon bunches containing at
least $10^{11-12}$ muons per bunch at collision. This implies
larger instantaneous stresses in the target.

  In order to design for the most challenging shock stresses at muon
colliders, the
modeling for this paper is benchmarked to the largest proton bunch charge
normally considered~\cite{status} at collision:
$4 \times 10^{12}$ muons per bunch; pion yield simulations for each
target are first
used to normalize the incident proton bunch charge to this capture rate.
On working backwards to the number
of pions and muons captured from the target, an assumed 25\% survival
rate~\cite{status} through the cooling channel and acceleration implies
initially capturing a total of $3.2 \times 10^{13}$ pions and muons,
where the two charge signs have been summed.

 As an aside, it is noted that this benchmarking procedure makes no
assumption on whether or not the pion capture and decay
channel is capable of capturing both charge signs in practice -- a
capability that seems plausible but has yet to be demonstrated in
muon collider design
studies -- because the same proton bunch charge will be required in
either case, and it is instead the bunch repetition rate that must
be doubled if only one pion sign is collected at a time.

  The bunch repetition rate is less critical than the proton bunch charge
vis-a-vis instantaneous shock stresses because, as section~\ref{sec:stress}
will show,
the shock waves die down quickly enough for the bunches to be relatively
independent in any reasonable muon collider bunch scenario.

  For a given proton bunch charge, the additional specification of
the bunch repetition rate determines the average proton beam power,
some fraction of which will be deposited as heat in the target band and will
need to be removed in the cooling tank. Proton beam powers of up to
7 MW~\cite{book96} have been assumed for some muon collider scenarios;
4 MW is a commonly-assumed value~\cite{status}. Constraints
on the muon beam currents from neutrino radiation may dictate less
powerful proton drivers for many-TeV muon colliders.
Design studies~\cite{study1,study2} for neutrino factories have
also assumed proton driver powers reaching the 1--4 MW level.

It will be seen in section~\ref{sec:cooling} that the target cooling
requirements are
rather relaxed even for such proton beam powers. This is due to the band
rotation spreading the heat load around the band circumference and to
the large band surface area exposed to the cooling water. Therefore,
the band target is unlikely in practice to set a limit on the average
proton beam power.

  Pion yield per proton is nearly proportional to proton energy, with
lower proton energies slightly preferred in the multi-GeV energy range;
equivalently, yield per MW of proton beam power falls slowly with
increasing proton energy. As a competing concern, higher proton energies
more easily enable the proton bunch lengths of 3 ns or less that are
optimal for a capture and decay channel that provides efficient capture
of the muons into rf acceleration while retaining substantial muon
polarization. We consider two representative proton energies, 6 GeV
and 24 GeV, in order to allow interpolation.

  For the band target design discussed here, the proton beam is incident
at a horizontal angle of 100 milliradians to the symmetry axis of the
solenoidal magnet capture channel (to maximize the pion yield) and is
focused to an elliptical beam spot at the target interaction region with
assumed gaussian profiles in both transverse dimensions and with r.m.s.
spot sizes tuned for acceptable shock heating stresses for the given
proton beam parameters and band material. The stress and yield simulations
assumed r.m.s. proton spot sizes of 2 mm (horizontal) and 15 mm (vertical)
incident on the Inconel and nickel bands, and 5 mm (horizontal) by 10 mm
(vertical) for the titanium alloy band.

\section{The Target Band}
%%%%%%%%%%%%%%%%%%%%%%%%%
\label{sec:band}

\begin{table*}[htb!]
\caption{Tabulation of some relevant properties of the candidate band
materials. The (range of) values for yield strength and fatigue strength were obtained
from the specified references.}
\begin{tabular}{|r|ccc|}
\hline
Property       &  Inconel  &  Ti-alloy & nickel \\ 
\hline
ave. atomic number, Z                  & 27.9 & 21.5 & 28.0 \\
ave. atomic weight, A                  & 59.6 & 46.8 & 58.7 \\
density ($\rho$), [${\rm g.cm^{-3}}$]  & 8.19 & 4.43 & 8.88 \\
interaction length ($\lambda$), [cm]   & 16.6 & 28.2 & 15.2 \\
radiation length (${\rm X_0}$), [cm]   & 1.55 & 3.56 & 1.48 \\
melting point,    [$^{\rm o}$C]        & 1298 & 1660 & 1450 \\
heat capacity, [${\rm J.K^{-1}.g^{-1}}$] & 0.435  & 0.526 & 0.46 \\
thermal conduct., [${\rm W.m^{-1}K^{-1}}$] & 11.4  & 6.7 & 60.7 \\
electrical conduct., [${\rm MS.m^{-1}}$]   & 0.8   & 0.56 & 14 \\
expansion coeff. ($\alpha$), [$10^{-5}/{\rm K}$] & 1.3  & 0.88 & 1.31 \\
elastic modulus (E), [$10^{11}\:{\rm N/m^2}$]     & 2.3  & 1.1 & 2.1 \\
0.2\% yield strength, [MPa]           &  1100~\cite{matweb}
                                      & $\sim 960$~\cite{matweb,TWA}
                                      & 59~\cite{matweb}  \\
fatigue strength [MPa], no. cycles       & 480-620 at $10^8$~\cite{NSref}
                                      & 510-700 at $10^7$~\cite{matweb,TWA} & N.A. \\
\hline
\end{tabular}
\label{band_materials}
\end{table*}

  The relevant properties of each of the 3 candidate target band
materials --  Inconel alloy 718,
titanium alloy 6Al-4V grade 5 and nickel --
are summarized in table~\ref{band_materials}.

  Inconel 718~\cite{matweb} is a
niobium-modified nickel-chromium-iron superalloy that was developed
for aerospace applications. Attractive properties include
high strength, outstanding weldability, resistance
to creep-rupture and resistance to corrosion from air and water.
It is used in high radiation
environments such as the core internals of light-water nuclear
reactors; examples of applications at accelerators include
high intensity proton beam windows and as the water-containment
material for proton beam degraders. It was proposed for beam windows
and for cladding the tungsten target elements
in the 170 MW proton beam at the Accelerator Tritium
Production (ATP) project (now part of the Advanced Accelerator Applications
initiative) and is the back-up candidate (behind 316LN
stainless steel) for the construction of Spallation Neutron Source
(SNS) target components.

 The elemental composition of Inconel alloy 718 that was used for pion
yield calculations is~\cite{matweb}
(with percentage by weight then molar fraction in the brackets):
Ni (54.3\%, 0.537),
Cr (19.0\%, 0.212),
Fe (17.0\%, 0.177),
Nb (5.1\%, 0.032),
Mo (3.1\%, 0.019),
Ti (0.9\%, 0.011),
Al (0.6\%, 0.013).

  The titanium alloy under consideration is titanium 6Al-4V (Grade 5),
consisting of titanium alloyed with 6\% aluminum and 4\% vanadium by
weight. This high-strength alpha-beta alloy is among the most versatile
and widely used of the titanium alloys. Applications include in pumps,
valves, turbines, aerospace and automotive parts, and vessels and casings
where corrosion is an issue. It offers ready
machinability and, unlike some alpha-beta titanium alloys, is not greatly
embrittled by welding. Titanium and titanium alloys have been used
in production targets, and this particular alloy was
recommended~\cite{Sieverscomm} after use in beam windows at CERN.

  Table~\ref{band_materials} shows nickel to have by far the lowest yield
strength and fatigue strength of
the three material options. However, nickel targets seem to evade
these low-strength predictions, with successful operation
in high power pulsed proton beams. For example, the currently
operating nickel target~\cite{pbar} at the Fermilab antiproton source has
absorbed peak energy depositions of up to 600\,J/g\ over 2.4 microseconds,
corresponding to an impressive $1100^o{\rm C}$ temperature rise.
It has been speculated that such nickel targets survive because they can
self-anneal in high power target environments, although the actual reason
for their exceptional performance is not well understood.

  As a concern for nickel targets, it was the experience of
both the FNAL anti-proton target~\cite{pbar} and BNL g-2 nickel
target~\cite{gminus2,pearsoncomm} that the nickel surfaces slowly
deteriorated and eventually began to powder on timespans of order
one year. The implications of this for the target replacement
lifetime and possible radioactive contamination would need to
be addressed for a muon collider target scenario.
Also, its much higher electrical conductivity will cause greater
drag on the target band from magnetic eddy currents in the capture
magnet. On the other hand, it will be seen that nickel's
pion yield is predicted to be slightly better than Inconel and
significantly better than titanium alloy, so it
may well be an attractive option for muon collider
scenarios with low repetition rate proton beam parameters
where the beam-induced damage will be minimized or eliminated
and the band can be rotated more slowly to reduce the
magnetic eddy currents.

  The requirements for a tightly focussed proton beam spot on the
target are more relaxed than, e.g., for an anti-proton production
target because any contribution from the spot size to the produced
pion beam emittance will tend to get washed out when the pions decay
to muons. The dimensions of the band webbing and proton beam spot
were chosen
to approximately maximize the pion yield while keeping the density
of energy depositions in the target to an acceptably low level.
General requirements for yield are that the proton path length
through the target material should be~\cite{status,Mokhovconf}
approximately 1.5--2 nuclear interaction lengths, and that the
band should be thin enough to allow most of the pions to escape
the target. It is predicted~\cite{book96,Mokhovconf} that high
atomic number (high-Z)
or medium-Z elements are favored over low-Z elements for the higher
end of the considered range of proton energies; this advantage is
less marked at lower proton energies. Inconel, titanium alloy and
nickel can all be considered to fall within the category of
medium-Z materials.

  Tilting targets by
approximately 100 milliradians with respect to the capture solenoid
has generally also been found~\cite{status,Mokhovconf}
to slightly increase the pion yield.
The elliptical beam spot was chosen solely to
reduce the beam-induced stress by spreading out the beam
energy deposition within the target.

\section{Target Band Drive and Support Mechanism}
%%%%%%%%%%%%%%%%%%%%%%%%%%%%%%%%%%%%%%%%%%%%%%%%%
\label{sec:drive}

  The target band rotates at of order 1 m/s, depending on the target
material and proton beam parameters, and with a rotation sense away from
the proton beam direction. Faster rotation minimizes heating pile-up from
successive proton pulses but the mechanical drive power must increase
as the square of the rotation velocity in order to compensate
for eddy current drag in the 20 tesla solenoid.

  As a numerical example of eddy current forces,
it can be roughly estimated that several hundred watts of
drive power would be required to overcome the eddy current forces 
from an Inconel band with the given cross-section and rotating at 1 m/s.
According to the ratio of electrical
conductivities in table~\ref{band_materials},
the eddy current power for this scenario would be be
$14/0.8=18$ times worse
if a nickel band was used instead of Inconel.

  The band is guided and driven by several sets of rollers located
around its circumference, as is shown in figure~\ref{layout}.
The motive power is most conveniently applied from those rollers within the maintenance
tunnel, where the radiation environment is less severe and maintenance
is easier. For most proton beam parameters, the eddy current drag will
not be large enough to require toothing the rollers and the parts of the
band they contact.
The tightest position tolerances on the rollers are the precisions of
1 mm or better required for the rollers defining the band's horizontal
position at interaction with the beam.

  Following the design of the BNL g-minus-2 target~\cite{gminus2}, the roller
assemblies will all incorporate self-lubricating graphalloy~\cite{graphalloy}
bushings. These commercially available bushings are manufactured from
molded graphite impregnated with metal and, in contrast to oil-based
lubricants, are compatible with high radiation environments.

\section{Considerations for Targetry in an Air Environment}
%%%%%%%%%%%%%%%%%%%%%%%%%%%%%%%%%%%%%%%%%%%%%%%%%%%%%%%%%%%
\label{sec:air}

  The pion production region
of the target is in an air environment.
This simplifies target maintenance and target band replacement
by avoiding any requirement to break and re-establish seals in
a high radiation environment.

  The vacuum window for the proton beam-line is located immediately
downstream from the final quadrupole magnet and a few meters upstream
from the production region. The proton beam spot size at this beam window
will be much larger than for the focused beam at the target interaction
region; this minimizes the peak beam-heating stresses and radiation damage
in the window and also simplifies the window cooling. The vacuum
in the pion decay channel begins at a beam window located (e.g.) 6 meters
downstream from the target interaction region.
These distances are not expected to
result in either excessive proton-air interactions upstream
from the target or significant degradation of the pion yield
since each meter of air corresponds to only
0.13 ${\rm g/cm^2}$ of matter, 0.14\% of an interaction
length, 0.33\% of a radiation length and to a minimum-ionizing
energy loss of only 0.24 MeV.

%..air circulation (done)
Following the procedure adopted for the BNL g-minus-2 target~\cite{gminus2},
activated air and gases from the target and interaction region are
continuously diluted and vented from the target hall into the outside
atmosphere. Initially, a loosely airtight
container around the target impedes gas transport away from the target
until most short-lived radio-isotopes have decayed. The iron plug
shown in figure~\ref{layout} may suffice for this purpose.
The activated air is then transported along the target hall to
allow dilution by mixing
with unactivated air until acceptable activation levels are reached for
venting into the outside atmosphere.

%  The cooling, installation and replacement of
%the band are discussed below in section~\ref{sec:band_cool}.
%Figure~\ref{layout} shows the water cooling tank and a band
%replacement work area inside a separate radiation-shielded maintenance
%area that surrounds the side of the band diametrically across
%from the target interaction region. This arrangement is expected
%to allow considerably easier access and maintenance on
%the band and the cooling tank.

\section{The Pion Capture and Decay Channel}
%%%%%%%%%%%%%%%%%%%%%%%%%%%%%%%%%%%%%%%%%%%%
\label{sec:channel}

%..air environment; beam window positions (done)

%..(done) magnets in capture/decay channel sim. to mercury option - list diff.

\begin{figure}[t!]
\centering
\includegraphics[width=4.0in]{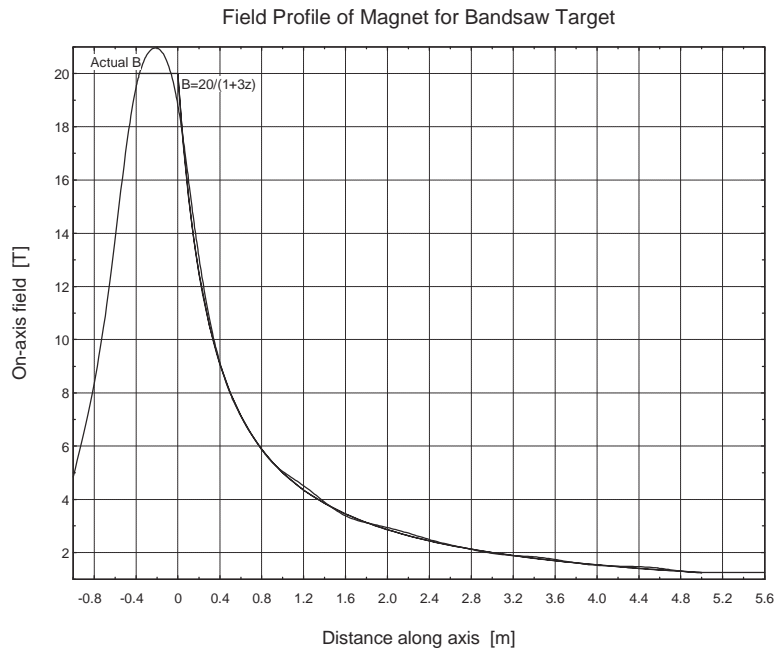}
\caption{The on-axis magnetic field profile in the solenoidal capture
channel. The plot shows, nearly superimposed, both the actual field
and the ``ideal'' field profile it was fitted to.}
\label{weggel_bfield}
\end{figure}

  The pion capture channel in figure~\ref{layout}
represents only a slight variation on channels considered
previously~\cite{status,study1,study2}.
The magnetic field in the solenoidal capture channel is nearly
identical to that in previous studies. As a minor change, no requirement
remains for field homogeneity upstream from the production region,
so no there is no constraint on how the upstream field rises to
the 20 tesla maximum.
On the other hand, the third coil block downstream from the upstream
end had to be moved outwards by approximately 10 centimeters to provide
adequate space for the band to exit the channel. A modest re-optimization
of the coil currents was required to restore the magnetic field map in
this region to the specifications of the previous studies.
The coil block positions and dimensions shown in figure~\ref{layout}
are taken directly from the computer programs used to optimize the
magnet geometry and magnetic field profile.
The re-optimized magnetic
field map is shown in figure~\ref{weggel_bfield}.

  The other requirement on the capture and decay channel that is
additional to previous scenarios is the provision of entry and exit
ports for the target band. The design of these ports is simplified by
the air environment of the
pion production region. The entry port need only
traverse the iron plug in the upstream end of the capture solenoid.
The downstream port is more challenging since it must traverse the
tungsten-based radiation shielding and then pass between the solenoidal magnet coil
blocks and out of the pion decay channel.

If it is considered undesirable to incorporate
such an exit port into a single cryostat, then
the alternative option exists of breaking the cryostat longitudinally into
two cryostats, so that the band can exit between them.
The exit port likely will
require some cladding with, e.g., tungsten carbide and water, in order
to shield the magnet coils
from any additional radiation load from low-energy neutrons.

   As is clear from figure~\ref{layout}, the target band exit port is
far enough upstream from the beam dump for it to be essentially irrelevant
in the beam dump design. Therefore, the beam dump design can be similar
to that of reference~\cite{beamdump}.

\section{Target Cooling}
%%%%%%%%%%%%%%%%%%%%%%%%
\label{sec:cooling}

\begin{figure}[t!]
\centering
\includegraphics[width=5.0in]{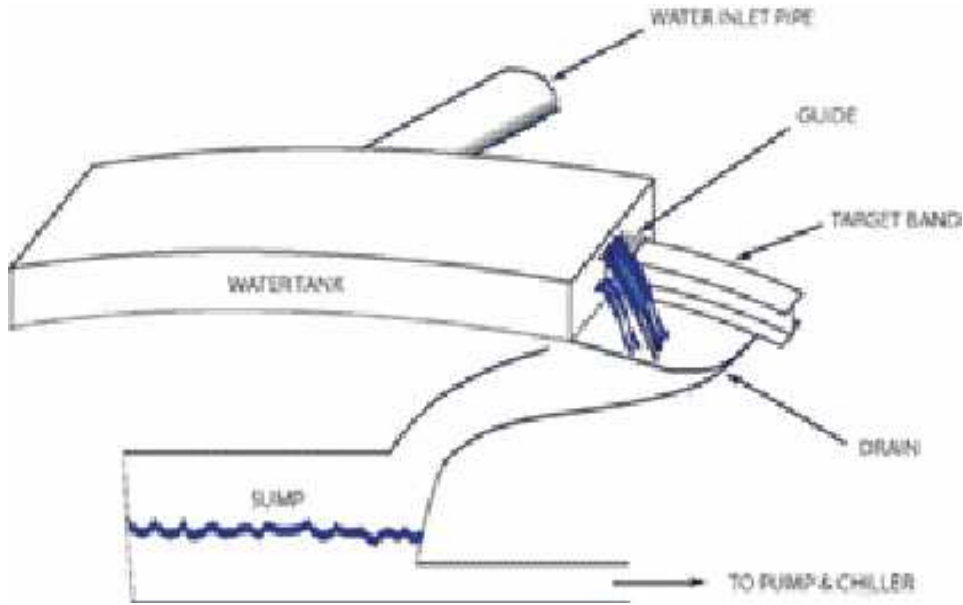}
\caption{A conceptual illustration of the target cooling setup.
A target band with an I-beam cross section is shown, as has been
assumed for the nickel and Inconel 718 material options. A
similar but simplified design would apply for the assumed
rectangular cross section of titanium alloy bands.
}
\label{band_cooling}
\end{figure}

  The heated portion of the band rotates through a 2 meter long
cooling tank whose conceptual design is shown in Fig.~\ref{band_cooling}.

  The water flows due to the gravitational head in a feeder tank,
with the band entrance
and exit ports in the ends of the tank also serving as the water outlets.
The flow rate can be simply adjusted by varying the head
in the feeder tank. Guides in the ports steer the water off to the
side of the target band and into a drain, to then be pumped through
a chiller and recirculated. The drains and structure at the ends of
the tank will be covered with hoods to prevent splashing 
(not shown in figure~\ref{band_cooling})
and, at the end where the band exits, high pressure air will blow
the residual water off the wetted band as it exits the hood.

  For equilibrium, the heat removed must balance the fraction of
the proton beam power deposited as heat in the band,
which MARS Monte Carlo computer
simulations found to be approximately 7\% (see section~\ref{sec:yield}),
i.e., approximately 70 kW
of heat deposited in the target band per megawatt of beam power.

  The 2 meter length of water in the cooling tank was chosen to be
sufficient to obviate the need for forced convection of the cooling
water for proton beam powers up to several MW, while many-MW proton
beam powers could be contemplated by incorporating forced convection
and/or increasing the cooling tank length. For the example of the I-beam
cross section for Inconel or nickel, the 0.69 square meters of immersed
target band surface area corresponds to an average heat transfer rate
of 10 ${\rm W/cm^2}$ per megawatt of beam power.
Even for proton beam powers up to several megawatts, this will
be comfortably below the 100 ${\rm W/cm^2}$
approximate maximum sustainable rate for
cooling by nucleate-boiling
with standing water under favorable conditions.
The cross section of the titanium band is 70\% as large, so the
heat transfer rates would need to be about 40\% higher.

  The water flow rate parameters are also relatively modest. For
example, an assumed 5 degree centigrade average temperature rise
in the water would require an exit flow rate of about 3.3 liters per
second per megawatt of incident beam power.
In the approximation that viscosity is neglected,
this flow rate could be
met by a combination of 1) a 2 m/s flow velocity supplied by pressure
from a 20 cm head of water and 2) an 18 ${\rm cm^2}$ cross-sectional
area per megawatt of beam power in each of
the 2 exit ports around the cross section of the target band.

 Concerning the desirability of drying the target between its multiple
passages through the cooling tank and subsequent exposures to the beam,
it is noted that this was not considered necessary for the BNL g-2
rotating-disk production target~\cite{gminus2}, which was simply left
wet. However, the motivation for air-drying is stronger for the geometry,
drive mechanism and larger local temperature rises of the rotating band
target considered here, and so it is assumed that drying air jets are
included in the design. As well as drying the bulk surface area, it
should be relatively straightforward to shape the air flow to also
remove all or almost all the water from the transverse gaps between
the 8 circumferential sections of the band and from the 3 circumferential
stress-barrier grooves at both the top and bottom of the webbing.

  As an attractive feature for maintenance, all equipment for the
cooling loop that requires moving parts -- 
the pumps, chiller, valves for the feeder tank, and air compressor --
can be conveniently located either inside the
maintenance tunnel or entirely outside the shielding walls surrounding
the target hall.

\section{Radiation Damage and Target Band Replacement}
%%%%%%%%%%%%%%%%%%%%%%%%%%%%%%%%%%%%%%%%%%%%%%%%%%%%%%
\label{sec:replacement}

\begin{figure}[t!]
\centering
\includegraphics[width=3.0in]{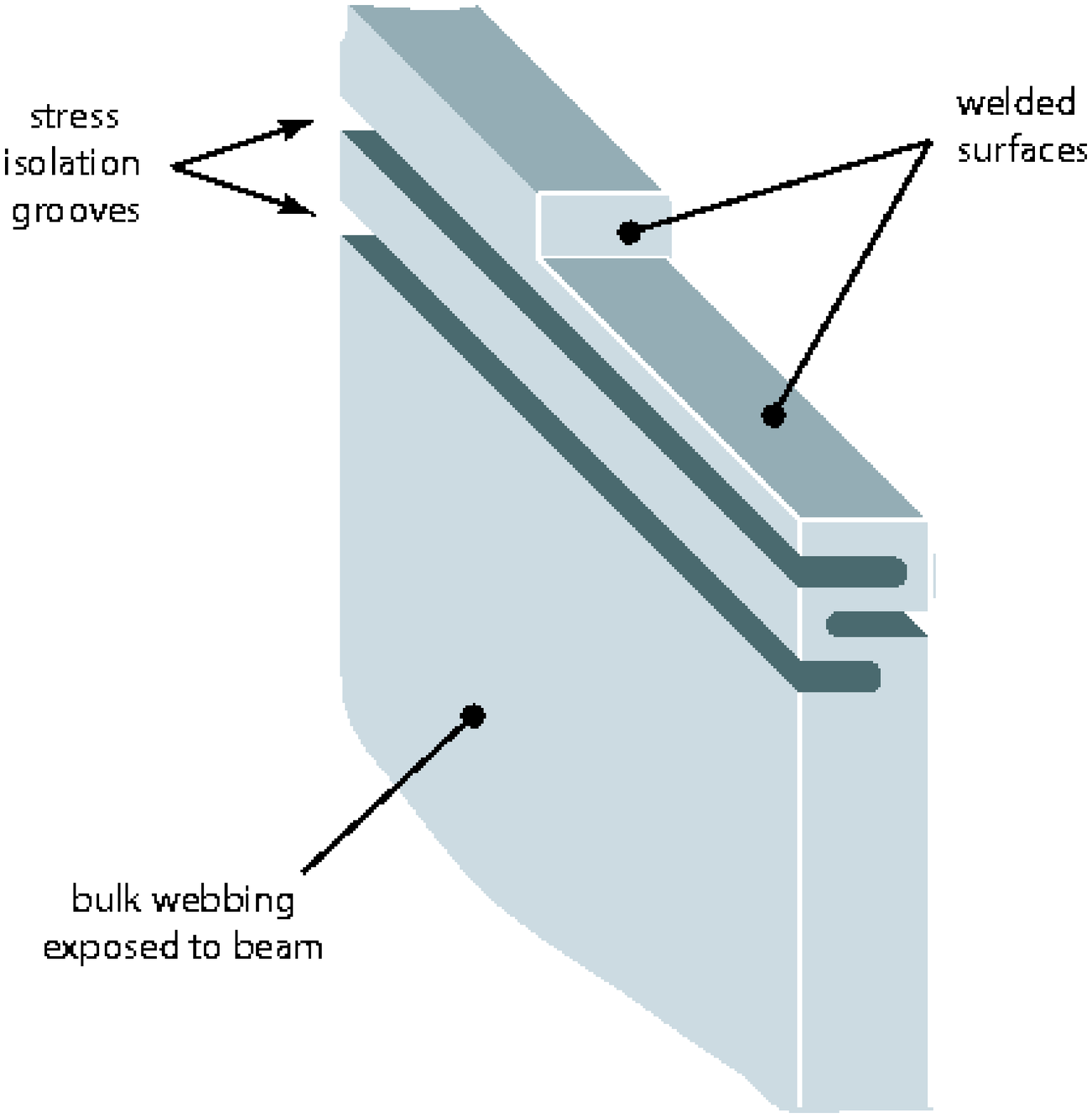}
\vskip 0.5in
\includegraphics[width=3.0in]{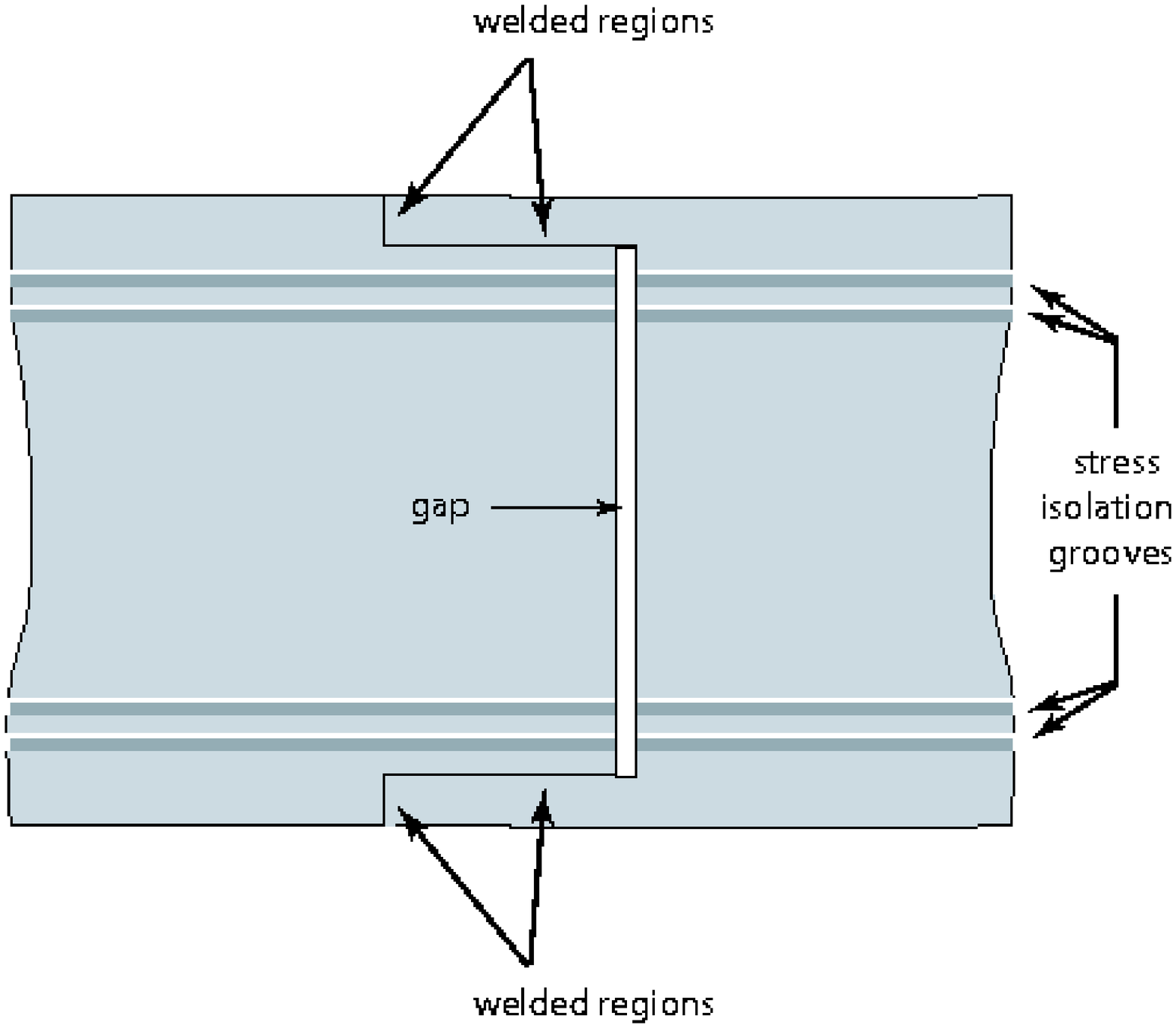}
\caption{
Three-dimensional view (top) of one corner of the end of one of the
eight circumferential segments that make up the band, and side view
(bottom) of a join between two of the segments. The figures illustrate
the use of grooves inside the welds to mechanically isolate them from
shock waves caused by the proton beam striking the mid-height region
of the webbing. The rectangular cross sectional geometry of the titanium
alloy band is shown; similar techniques can be used to isolate the welds
for the I-beam cross section of the Inconel or nickel bands.}
\label{bandweld}
\end{figure}

  The rotation of the target band has the desirable dilution effect
that the rate of radiation damage on any particular section
of the band material is reduced by roughly two orders of magnitude
relative to a fixed target geometry, because the region of maximum
energy deposition from any particular proton bunch has a characteristic
width on the order of one interaction length (i.e. 15--28 cm), whereas
the 15.7 meter band circumference corresponds to 55--100 interaction
lengths, depending on the target material.
Even so, the strength and other mechanical
properties of the target band will likely eventually be degraded
by repeated shock heating stresses and radiation damage to the point
where the band needs to be replaced. Therefore, the target design must
allow for the routine removal and replacement of the target band. 

  A very approximate determination of radiation damage to the target
band can be obtained from the estimated fluence of particles through
the target material and the rule-of-thumb that 1 displacement per
atom (dpa) will be produced by a fluence $10^{21}$ minimum ionizing
particles per square centimeter. This predicts that a few-MW proton
beam would produce of order 1 displacement per atom (dpa) per year of
radiation damage. In turn, this suggests that annual replacement of
the band should easily suffice even for the highest power proton
beams under consideration since, for comparison, a 6 dpa
design lifetime has been set for the 316LN steel (or Inconel 718,
as a back-up) target components in the SNS.

  Welds can be a potential Achilles heel for high-stress targetry
applications. Favorable features for the rotating band target
geometry in this regard are that no welds are required between
dissimilar metals and that the welds can be
placed at the top and bottom of the band webbing, away from the
mid-height region that receives the beam energy. For further protection,
circumferential grooves placed inside the welds can mechanically
isolate them from shock waves emanating from the beam interactions.
A welding scenario incorporating such grooves is shown in
figure~\ref{bandweld}.

  With three grooves inside each weld, as shown in figure~\ref{bandweld},
the shock waves emanating from the target
region will be almost entirely reflected back into
the central region or else dissipated by multiple scatters. This
will effectively shield the weld from the shock-heating transients.

  Because the band is not load-bearing, each of the grooves can extend
nearly through the thickness of the webbing, and they are assumed to
run all the way around the band circumference. The removal of material
away from the mid-height production region should be irrelevant, or
perhaps even slightly beneficial, to the pion yield.

  The join region can be blown dry after exiting the cooling tank.
A few-millimeter gap might well be retained in the joins between
the 8 circumferential band segments, in order to avoid leaving
small cracks that could retain water by capillary action
after passage through the cooling tank.
Such gaps will have a negligible effect on the pion yield for those
proton pulses passing through the join regions, since they represent
only of order a one percent reduction in the effective target length.
The yield is insensitive to such small changes
near the optimal beam intersection length because,
by definition, the optimal beam intersection length for yield
occurs where the first derivative of yield with respect to length
is zero.

  Each of the candidate band materials is suitable for welding.
Inconel 718 gives outstanding weldability~\cite{matweb} and
resistance to post-weld cracking. Ti-6Al-4V is among
the better alpha-beta titanium alloys for welding~\cite{inox}
and is weldable in the annealed condition as well as in the
solution-treated and partially-aged conditions.

%..radiation damage calculation
%..takes place in maintenance tunnel
%..hot band should be removed robotically:
%....cutting band piece by piece, in 1 m lengths
%....clamp band around cut and guillotine
%....drop into hot box
%....close and remove hot box - now can access tunnel
%..installing band:
%....are welding pre-shaped lengths of e.g. 2m
%....weld is on flanges of I-beam x-section - away from stress
%....a small gap in webbing should be OK (but see stress section)

%..maintenance tunnel - pointer to section on band installation

  Target bands will be installed and extracted from the dedicated
band maintenance area located in the maintenance tunnel
(see figure~\ref{layout}). Remote extraction is presumably the only
viable option for heavily irradiated used bands. The band will
be removed from its channel by progressively clamping and
then shearing off (e.g.) 1 meter lengths and dropping them into a hot box.
It is expected that, once the hot box has been locked shut and the
irradiated band removed to a disposal area, radiation levels in the
maintenance tunnel will have fallen to an acceptably low level to
allow the immediate manual installation of the new band without the
need for a cool-down period. This assumption should eventually
be checked using particle tracking simulations (e.g. with MARS~\cite{MARS})
that can determine the level of residual radiation carried into the
maintenance area by the target band and by neutrons leaking through
the band ports in the shielding wall, although these levels are expected
to be similar to those calculated in reference~\cite{beamdump}.

 In what is almost the reverse procedure to band removal, the
new band will be progressively welded together in situ from (e.g.)
eight 1.96 meter long chords of target band that have been previously
formed into the correct cross section and circumferential curvature.

\section{MARS Monte Carlo Simulations of Pion Yield and Beam Energy Deposition}
%%%%%%%%%%%%%%%%%%%%%%%%%%%%%%%%%%%%%%%%%%%%%%%%%%%%%%%%%%%%%%
\label{sec:yield}

\begin{table*}[htb!]
\caption{A summary of MARS and ANSYS predictions for pion yields,
energy depositions and stresses. Units are indicated in square brackets.
The superscript ``3.2'' refers to the proton bunch charge that results
in a total of $3.2 \times 10^{13}$ captured pions.
See text for further definitions and details.}
\begin{tabular}{|r|cc|cc|cc|}
\hline
{\bf band material} &
\multicolumn{2}{|c|}{ {\bf Inconel 718}} &
\multicolumn{2}{|c|}{ {\bf Ti-alloy}} &
\multicolumn{2}{|c|}{ {\bf nickel}} \\ 
proton energy [GeV]              &  6     &  24      &  6     &  24      &  6     &  24    \\
\hline
captured $\pi^+$ yield/proton    & 0.102  & 0.303    & 0.080  & 0.249    & 0.102  & 0.302  \\
captured $\pi^-$ yield/proton    & 0.105  & 0.273    & 0.083  & 0.224    & 0.105  & 0.292  \\
$ppp^{3.2}$  $[10^{13}]$         & 15.5   & 5.56     & 19.6   & 6.78     & 15.5   & 5.39   \\
$E^{3.2}_{pulse}$   $[kJ]$       & 149    & 214      & 188    & 260      & 149    & 207    \\
$U^{3.2}_{max}$     [J/g]        & 32.0   & 31.7     & 25.6   & 21.3     & 32.5   & 37.4   \\
$\Delta T^{3.2}_{max}$
                   [${\rm ^oC}$] &  74    &  73      &  49    &  40      &  71    &  81    \\
stress, $VM^{3.2}_{max}$
              [MPa]              & 330    & 360      &     72   &   68   & 330    & 340    \\
\% of fatigue strength           & 53-69\%&58-75\%   & 10-14\%  & 10-13\%& N.A.   & N.A.   \\
\hline
\end{tabular}
\label{tab:predictions}
\end{table*}

 Full MARS~\cite{MARS} tracking and showering Monte Carlo simulations
were conducted for 6 GeV and 24 GeV protons incident on the
target, returning predictions for the pion yield and energy deposition
densities.

\begin{figure}[t!]
\centering
\includegraphics[width=3.5in]{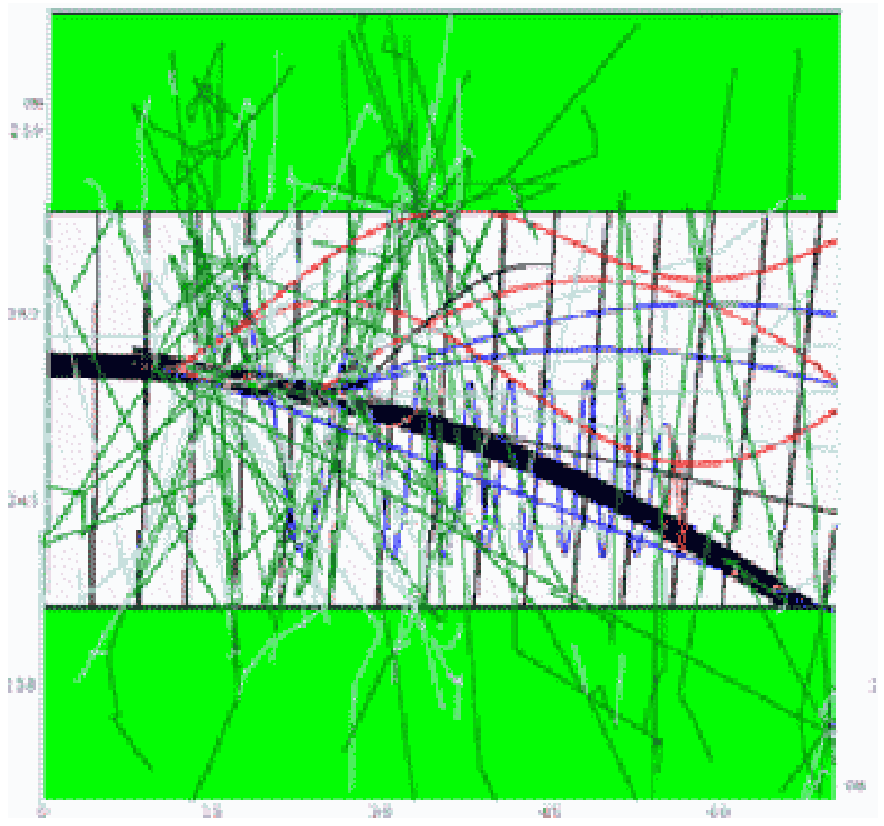}
\caption{MARS Monte Carlo simulation of secondary particle production
from 5 interactions of 24 GeV protons in an Inconel band target.
}
\label{marsgeom}
\end{figure}

\begin{figure}[t!]
\centering
\includegraphics[width=3.5in]{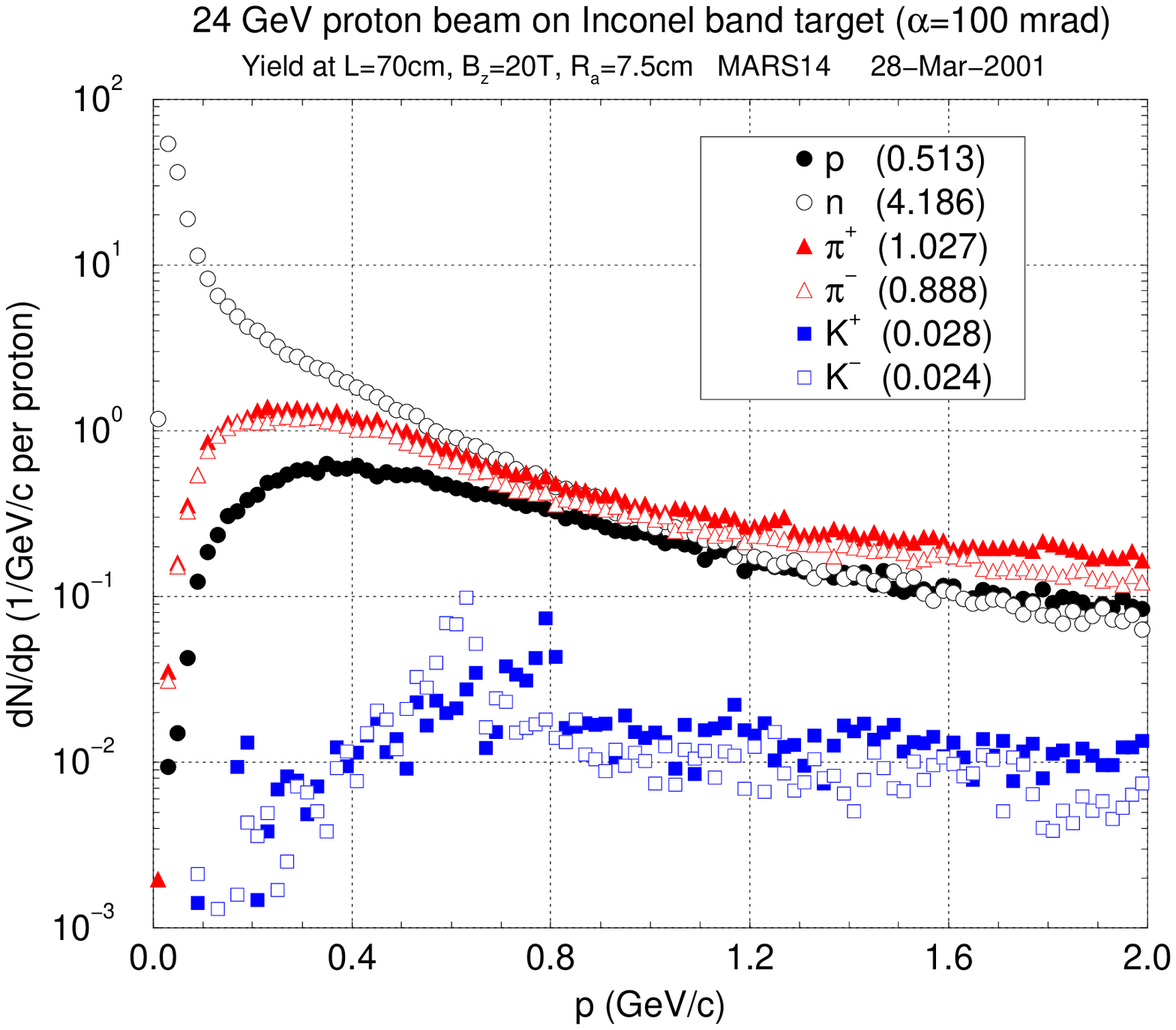}
\caption{
Momentum distribution of hadron yields for
24 GeV protons interacting in an Inconel band target.
}
\label{phadron}
\end{figure}

\begin{figure}[t!]
\centering
\includegraphics[width=3.5in]{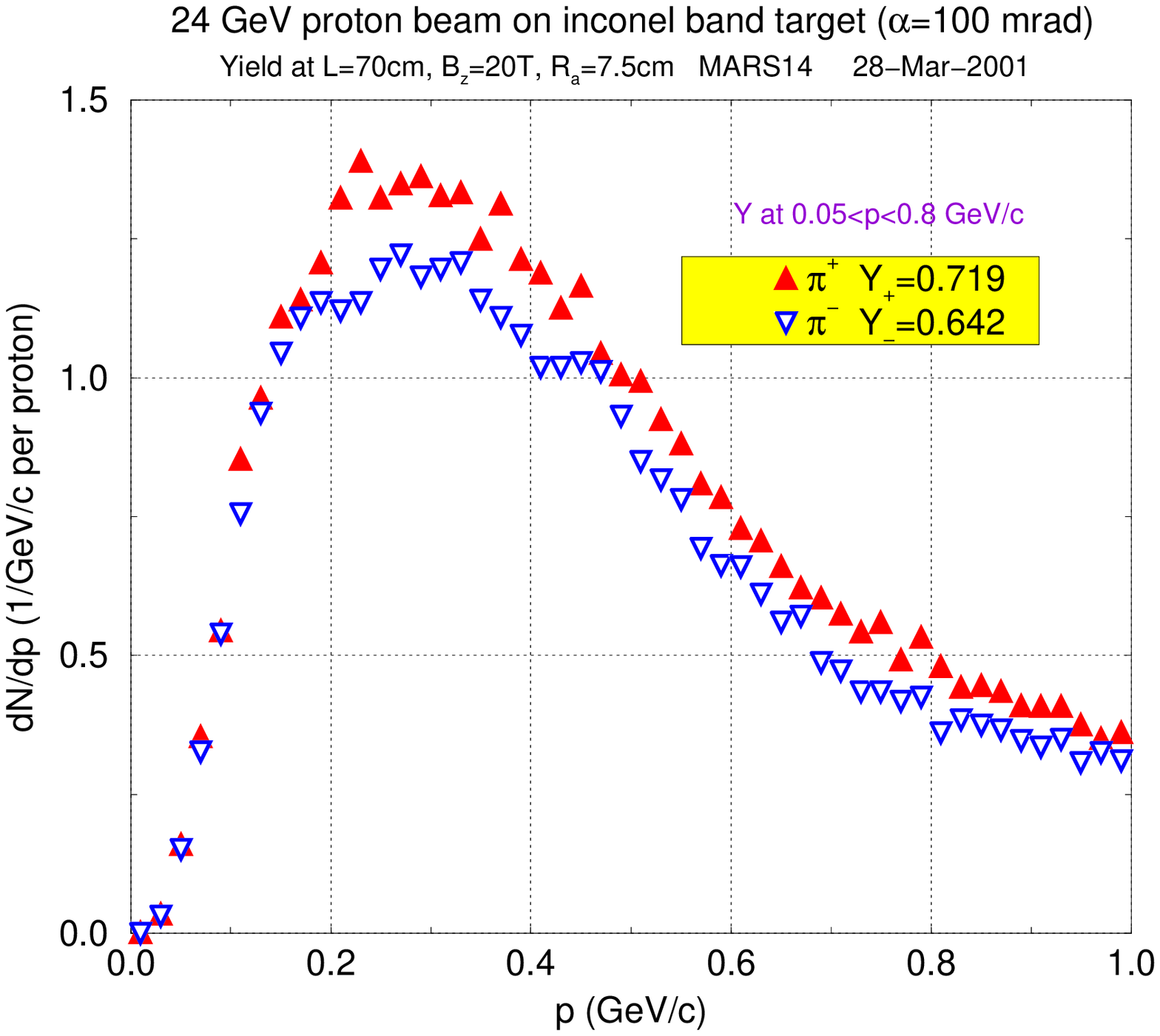}
\caption{
Momentum distribution of pion yields for
24 GeV protons interacting in an Inconel band target.
}
\label{ppion}
\end{figure}

\begin{figure}[t!]
\centering
\includegraphics[width=3.5in]{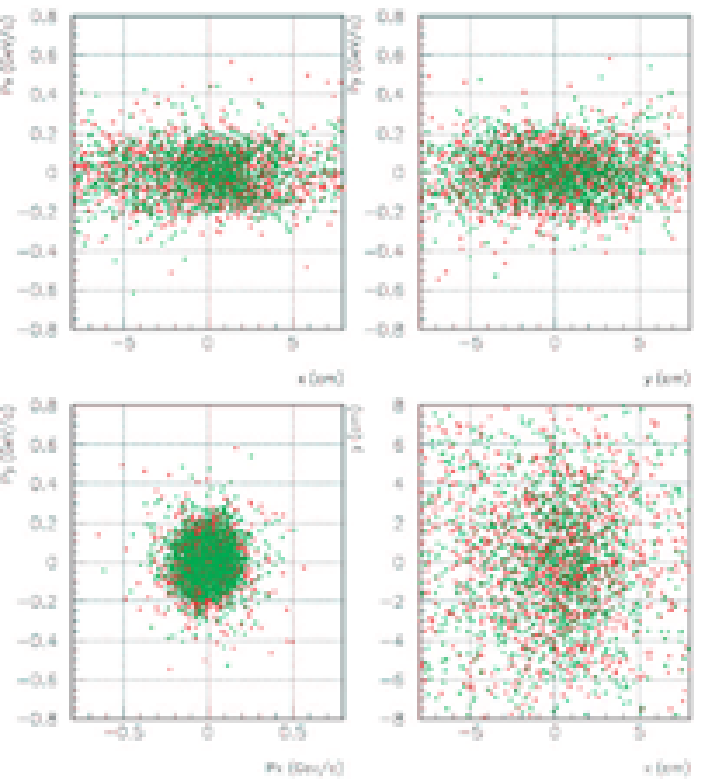}
\caption{
Phase space distributions of pions produced from
24 GeV protons interacting in an Inconel band target.
Shown are
(i) x-component of momentum vs. x position (top left),
(ii) y-component of momentum vs. y position (top right),
(iii) y vs. x components of momentum (bottom left) and
(iv) y vs. x position components (bottom right).
}
\label{scatter}
\end{figure}

  The detailed level of the MARS simulations is illustrated by
figure~\ref{marsgeom},
using the example of several 24 GeV proton interactions in an Inconel band.
Figure~\ref{phadron} shows the corresponding yield and momentum spectra
for all hadrons; figure~\ref{ppion} gives more detailed
information for the pions. Several scatter plots to illustrate
the distribution in phase space of the produced pions
are displayed in figure~\ref{scatter}. The plots are seen to be
relatively symmetric in the x and y coordinates, which indicates
that any asymmetries due to the band tilt and elliptical beam
spot are largely washed out by the large phase space volume
occupied by the produced pions.

 The yield per proton for positive and negative pions-plus-kaons-plus-muons
at 70 cm
downstream from the central intersection of the beam with the target
was predicted for the kinetic energy
range 32$<E_{kin}<$232~MeV that approximates the capture
acceptance of the entire cooling channel. Note that the material
in the flanges of the I-beam for the Inconel and nickel targets
was not included in the calculation;
its inclusion might result in a small change in the predicted
yield.

  Table~\ref{tab:predictions} summarizes the yield and energy deposition results
from the MARS calculations. It includes the several rows of derived results that
assume the scenario, taken from section~\ref{sec:protons}, of $3.2 \times 10^{13}$
captured pions. These derived quantities are identified with a superscript ``3.2''
and include: the required number of protons per pulse, $ppp^{3.2}$, the
required total proton pulse energy, $E^{3.2}_{pulse}$, the maximum localized
energy deposition in the target material and corresponding temperature
rise, $U^{3.2}_{max}$ and $\Delta T^{3.2}_{max}$.

  Approximately 7\% of the proton beam energy is deposited in the
target. Detailed 3-dimensional maps of energy deposition densities
were generated for input to the dynamic target stress calculations
that are discussed in the following section.

%[Nikolai, do you wish to do these?]
%  The peak energy deposition density was found to be ??~J/g per
%pulse, corresponding to a temperature rise of $\Delta T$=??$^{\circ}$C
%and a total power dissipation in the target of 0.???~MW.
%Contributions to the deposited energy come from
%dE/dx from hadrons and muons (??\%), electromagnetic showering (??\%) and
%from absorbtion of sub-threshold particles (??\%).
%Power dissipation in the inner layer of tungsten shielding (7.5$<$r$<$15~cm)
%was also determined, and was found to be
%0.???~MW.

\section{Shock Heating Stresses}
%%%%%%%%%%%%%%%%%%%%%%%%%%%%%%%%
\label{sec:stress}

\begin{figure}[t!] % fig 1
\centering
\includegraphics[height=2.5in]{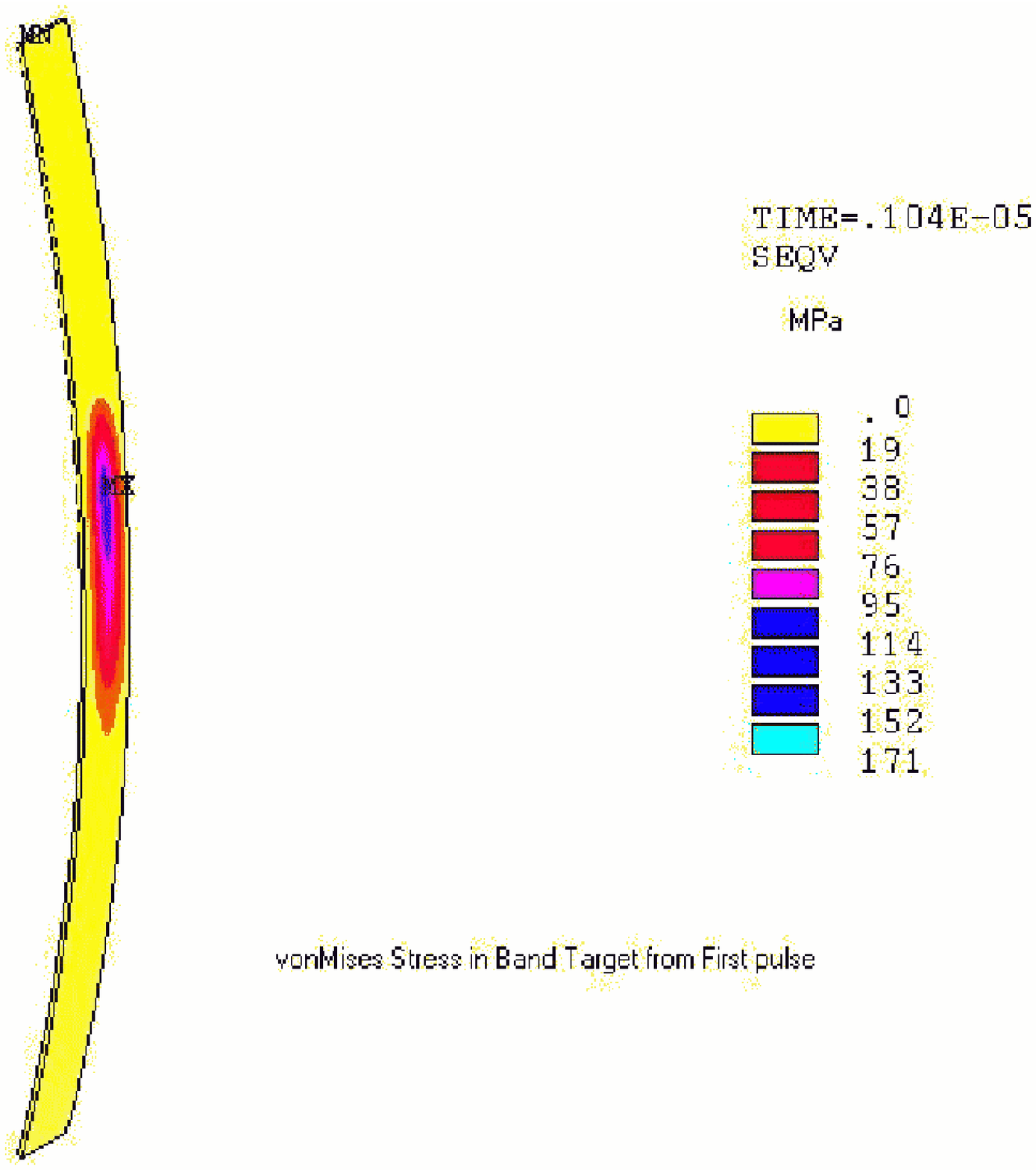}
\caption{
Predicted
von Mises stress distribution for an Inconel target band
at one microsecond after exposure
to an instantaneous proton bunch of
$1.7 \times 10^{13}$ 24 GeV protons. This is a smaller bunch charge
than would be typical for muon colliders; the distribution of stress values
will scale in approximate proportion to the bunch charge unless the
material's fatigue strength is exceeded.
%Figure reproduced from reference
%~\cite{study2}.
}
\label{band_vonMises_initial}
\end{figure}

\begin{figure}[t!]
\centering
\includegraphics[width=3.5in,angle=270]{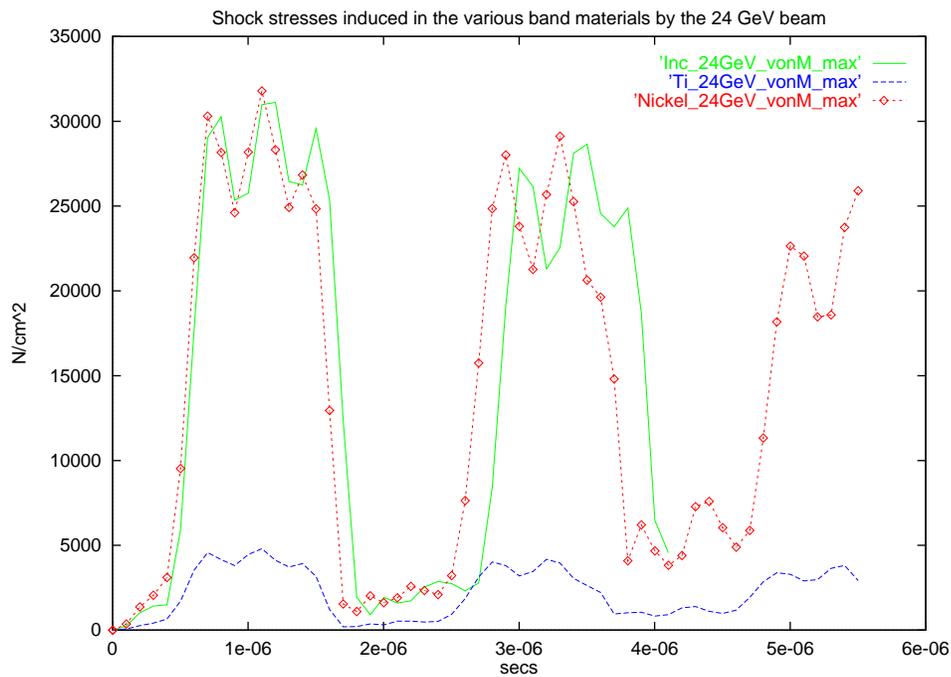}
\caption{
Predicted time dependence of von Mises stresses on Inconel 718, titanium alloy
and nickel bands due an instantaneous energy deposition from a bunch
of $1.5 \times 10^{14}$ 6 GeV protons with transverse dimensions as given in
table~\ref{target_band_specs}.
The time origin
corresponds to the arrival of the proton pulse. The stress values
are shown for the position of maximum stress in all cases.
}
\label{nick4}
\end{figure}

\begin{figure}[t!]
\centering
\includegraphics[width=3.5in,angle=270]{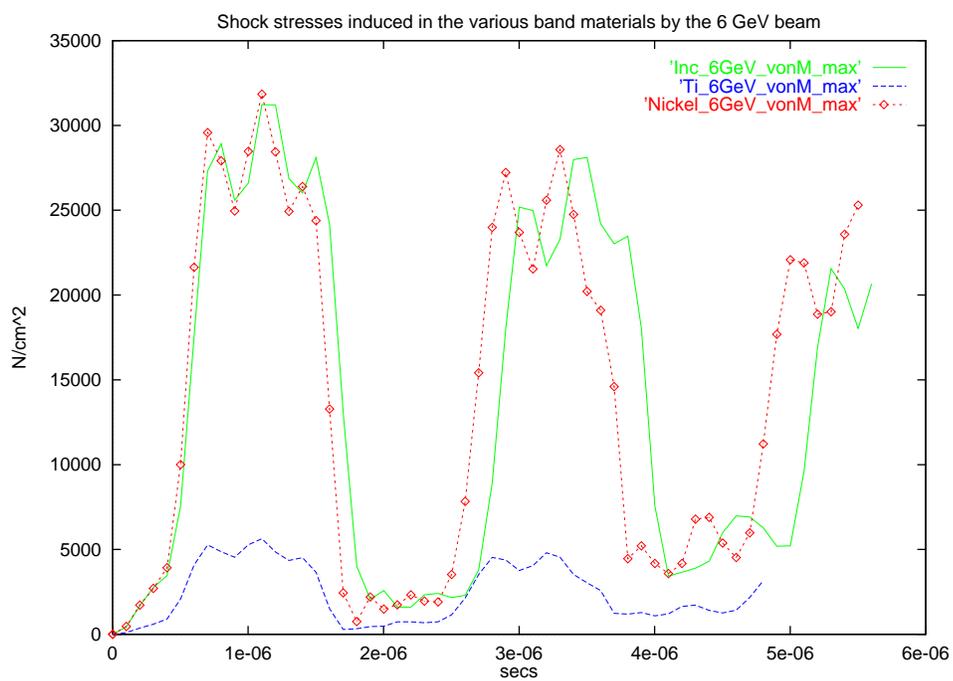}
\caption{
Similar to figure~\ref{nick4}, but for an incident bunch
of $5 \times 10^{13}$ 24 GeV protons.
}
\label{nick3}
\end{figure}

\begin{figure}[t!]
\centering
\includegraphics[width=3.5in,angle=270]{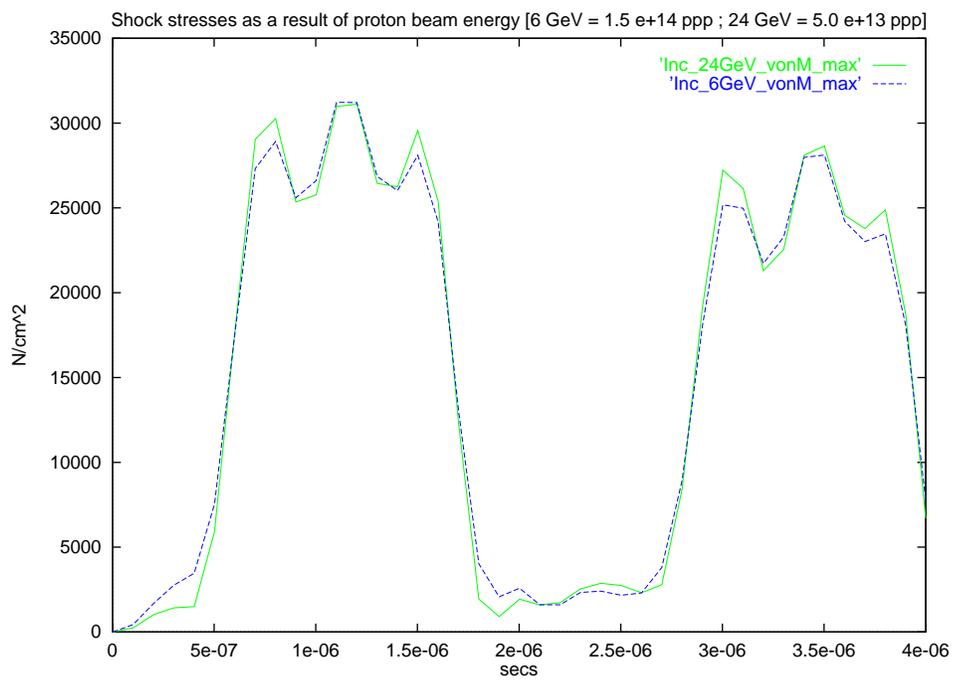}
\caption{
The curves for the Inconel target from figures~\ref{nick4}
and~\ref{nick3}, for 6 GeV and 24 GeV proton beams respectively,
showing the close correspondence in the stress time
development.
}
\label{nick7}
\end{figure}

\begin{figure}[t!]
\centering
\includegraphics[width=3.5in,angle=270]{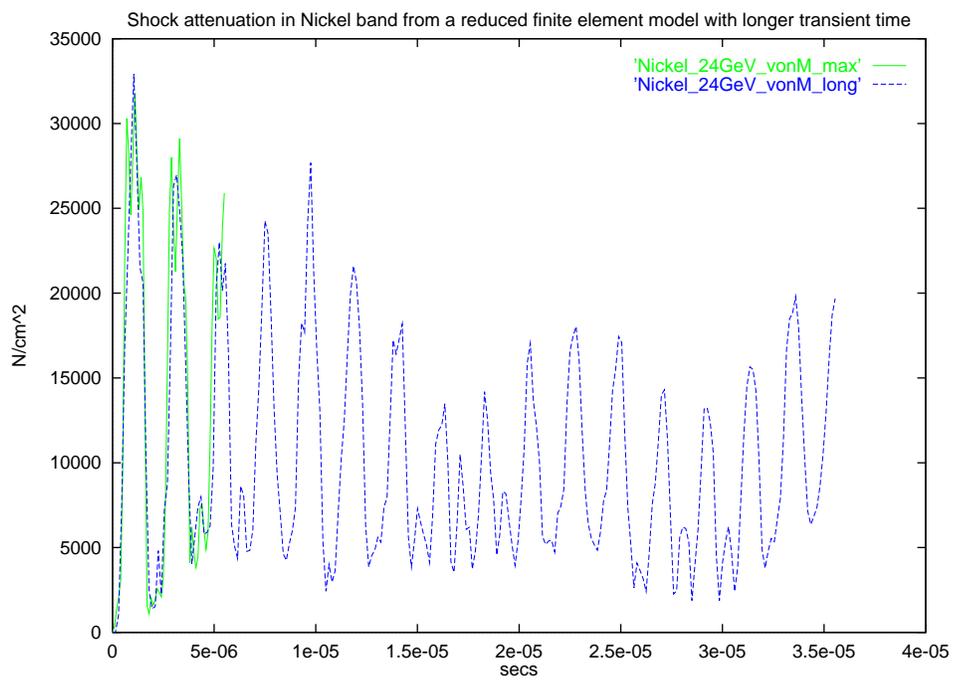}
\caption{
Similar to figure~\ref{nick3}, for $5 \times 10^{13}$ 24 GeV protons on a nickel
target, but extended to larger time values to show the dissipation of the shock stresses
after multiple reflections from the band surfaces.
}
\label{nicknew}
\end{figure}

\begin{figure}[t!]
\centering
\includegraphics[width=3.5in,angle=270]{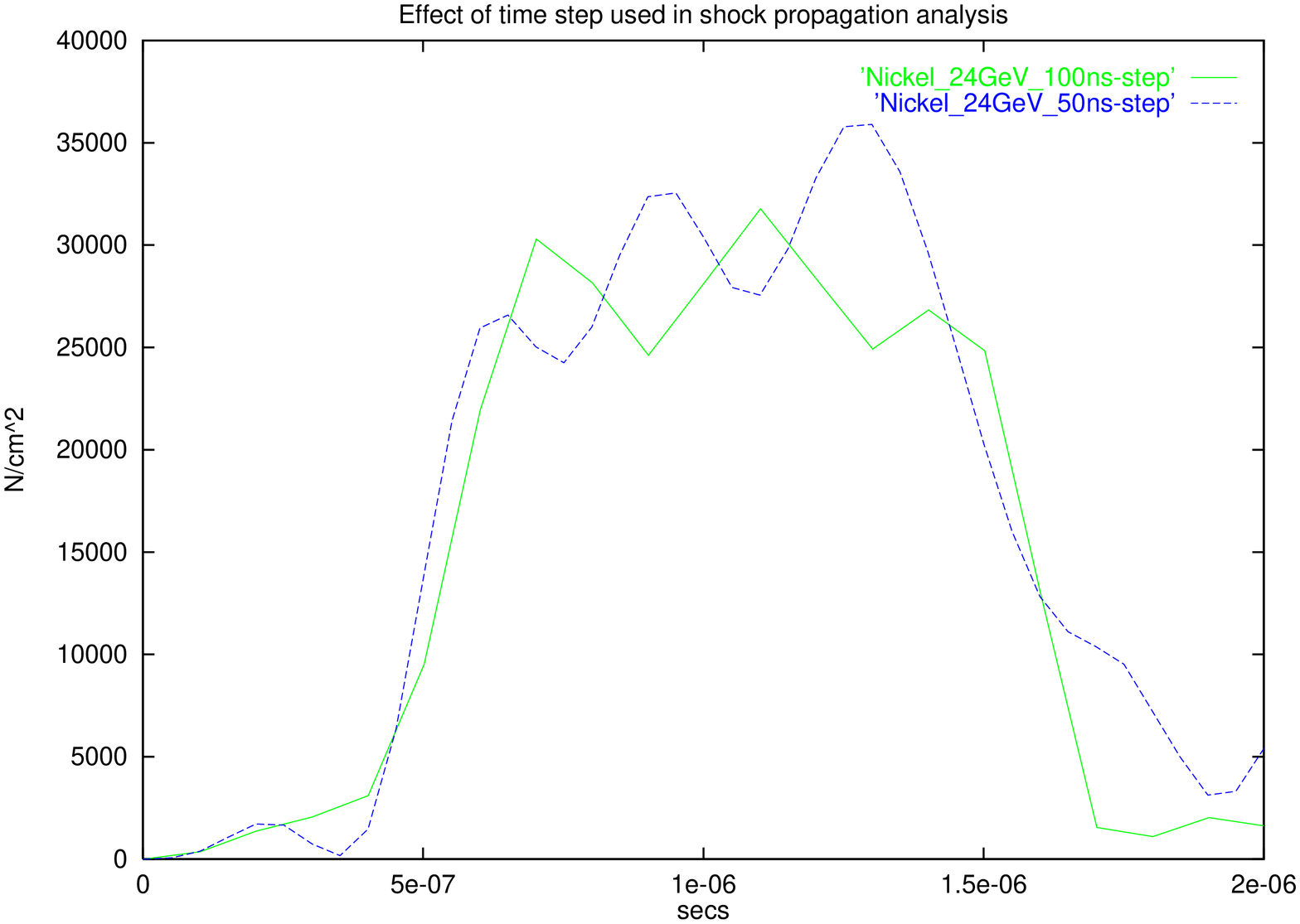}
\caption{
Similar to figure~\ref{nick3}, for $5 \times 10^{13}$ 24 GeV protons on a
nickel target but for both 50 ns and 100 ns time steps in the ANSYS simulation.
The reasonable
agreement between the two curves suggests that the normal 100 ns step size is
adequately short for approximate stress predictions.
}
\label{nick6}
\end{figure}

  Probably the most critical issue faced in solid-target design scenarios
for pion production at neutrino factories or muon colliders is the
survivability and long-term structural integrity of solid targets in
the face of repeated shock heating. To investigate this, finite element
computer simulations of the shock heating stresses have been conducted
using ANSYS, a commercial package that is widely used for stress
and thermal calculations.

  The target band geometry was discretised into a 3-dimensional
mesh containing approximately 30 000 elements. This was as fine
as the computing capacity and memory allowed and was judged to be
adequate for the accurate modeling of shock wave propagation.

 The ANSYS simulations conservatively assumed that the deposited energy
is all converted to an instantaneous local temperature rise.
The dynamic stress analyses were preceded by a transient thermal
analysis to generate temperature profiles using as input the 3-dimensional
energy deposition profiles previously generated by MARS for the
production assumption of $3.2 \times 10^{13}$ total captured pions (see the
preceding section).

 Dynamic stress calculations were then performed
both for a ``free edge'' band, i.e., with no I-beam flanges, and with
a ``fixed edge'' constraint, in which the edges of the band are constrained
against displacement in both the radial and axial direction. The ``free edge''
boundary condition is appropriate for the titanium alloy band;
the ``fixed
edge'' model is considered likely to provide
an improved approximation
to the Inconel and nickel bands with their I-beam flanges
without requiring the extra computing
capacity that would be needed to simulate the more complicated true geometry.

 The von Mises stress (i.e., the deviation from the hydrostatic state
of stress) was found to be initially zero but to develop and fluctuate
over time as the directional stresses relax or are reflected from material
boundaries. Figure~\ref{band_vonMises_initial} gives an example snapshot
of the
predicted von Mises stress distribution at one microsecond after the arrival
of a proton pulse, and the remaining figures~\ref{nick4} to~\ref{nick6}
show various aspects of the predicted stress at the position of maximum stress,
respectively: the time development for 6 GeV protons and for all three band
material candidates; the same for 24 GeV protons; superimposed plots for
6 GeV and 24 GeV protons and for the nickel band; the stress development over
a long enough time-span to see the attenuation
of the stress levels; and a check
on the time step used in the ANSYS calculations.

  Table~\ref{tab:predictions} summarizes the ANSYS predictions for
the maximum stress created at any time and any position in each of the band
materials, $VM^{3.2}_{max}$. These values were obtained by
reading off from figures~\ref{nick4} and~\ref{nick3} and then scaling
to the bunch charge for a total yield of $3.2 \times 10^{13}$ captured pions.
The final row of table~\ref{tab:predictions} displays the percentage of the
fatigue strength (from table~\ref{band_materials}) that this represents.

  For the Inconel band, the calculated fraction of the fatigue strength that
the band would be exposed to in this ``worst case'' proton bunch scenario,
53-69\%, is either close to or slightly above
what could be considered a safe operating margin for the target band.
A more definitive determination of the proton beam parameters that allow
survivability and adequate safety margins for this target
scenario could be provided by data from the ongoing BNL E951 targetry
experiment~\cite{E951}, with planned stress tests for bunched 24 GeV
proton beams incident on several types of targets, including Inconel 718.
The Inconel target may well be appropriate for some proton beam specifications
at a muon collider, and it has already been shown~\cite{study2} to likely
give a wide safety margin for the more relaxed beam parameters of neutrino
factories.

  The titanium alloy was predicted to have a very conservative safety margin
even for the assumed muon collider beam parameters: only 10-14\% of the fatigue
strength. Although the yield is about 20\% lower than for the other two
candidate materials, target bands from titanium alloys look likely to survive
with any proton bunch charges that might reasonably be contemplated for muon
colliders.

  Finally, nickel targets are known to evade the predictions for fatigue
strength limits, as already mentioned. Test beam experiments would be
required to establish the suitability or otherwise of a nickel band
production target for any particular muon collider scenario.

 All of the above calculations apply for a circumferentially continuous
band. It remains to check the level of von Mises stresses
at the gaps  between the eight welded band sections, although
it is noted that the BNL g-2 target was
deliberately segmented longitudinally in order to reduce the beam
stresses. For rotating band targets in muon colliders, additional
periodic slots in the webbing may also be considered for thermal
stress relief and eddy current reduction in rotating band targets
for muon colliders.

\section{Conclusions}
%%%%%%%%%%%%%%%%%
\label{sec:concl}

 In summary, the Inconel rotating band target design appears to
be a promising option for pion production targets at muon colliders.
The design concept appears to be manageable from an engineering point
of view, and initial simulations of
target yields and stresses are encouraging for each of three candidate
target materials: Inconel 718, titanium alloy 6Al-4V grade 5 and nickel.

 Priorities for further evaluation of this target scenario include
engineering designs of the components, optimization of the
band geometry for pion yield and calibration of the target stress
predictions to experimental targetry results.

\section{Acknowledgements}
%%%%%%%%%%%%%%%%%%%%%%%%%%

  We acknowledge helpful discussions with Charles Finfrock, George
Greene and Charles Pearson, all of BNL.  This work was performed under
the auspices of the U.S. Department of Energy under contract no.
DE-AC02-98CH10886.


\begin{thebibliography}{99}
\bibitem{graphalloy}  Graphalloys are manufactured from molded graphite
impregnated with metal. Graphalloy is a registered trademark
of the Graphite Metallizing Corporation.
\end{thebibliography}
\end{document}